\documentclass{PoS}

\usepackage{epsfig}
\usepackage{bm}

\title{Universal chemical freeze-out as a phase transition signature}

\ShortTitle{Universal chemical freeze-out as a phase transition signature}

\author{\speaker{Ulrich Heinz} 
\thanks{Work supported by the U.S. Department of Energy,
        grant DE-FG02-01ER41190.}\ \  and Gregory Kestin\\
        Department of Physics, The Ohio State University, 
        Columbus, OH 43210, USA\\
        E-mail: \email{heinz@mps.ohio-state.edu}}

\abstract{It is shown that kinetic freeze-out in relativistic heavy-ion
collisions invariably entails a non-trivial dependence of the freeze-out
temperature on the collision centrality. The centrality independence of 
the chemical freeze-out temperature observed in Au+Au collisions at RHIC 
is therefore inconsistent with the hypothesis that hadron abundances 
decouple kinetically from inelastic hadron-hadron interactions. On the 
other hand, it is consistent with the hypothesis that chemical decoupling
is driven by the quark-hadron phase transition, and that the observed
universal chemical freeze-out reflects its critical temperature, 
independent of the dynamical state of the collision fireball as it 
passes through the phase transition.}
  
\FullConference{Critical Point and Onset of Deconfinement\\
                July 3-6, 2006\\
                Galileo Galilei Institute, Florence, Italy}

\begin{document}

\section{Introduction}

From Lattice QCD \cite{KL03} we know that strongly interacting matter
of zero net baryon density undergoes a deconfining and chiral symmetry 
restoring phase transition from a hadron resonance gas at low temperatures 
to a quark-gluon plasma (QGP) at high temperature. The critical temperature 
of this transition has for several years been quoted \cite{KL03} as 
$T_c = 173\pm15$\,MeV (with an associated critical energy density 
$e_c\approx 0.7$\,GeV/fm$^3$), but recently this value has been challenged. 
Since (at least for systems with small net baryon density such as those 
created in heavy-ion collisions at Relativistic Heavy Ion Collider (RHIC) 
energies near midrapidity) the phase transition is not a sharp singularity 
but rather a continuous cross-over where thermodynamic quantities change 
dramatically over a relatively narrow temperature range \cite{Fodor_Nature}, 
the definition of $T_c$ is not unique, and one obtains slightly different
values from different types of observables \cite{Fodor_Tc}. Depending
on the choice of Lattice QCD action and extrapolation to the continuum,
different groups have recently obtained $T_c$ values ranging from 
$T_c(\chi_{\bar\psi\psi})=151\pm3\pm3$\,MeV (extracted from the peak
in the chiral susceptibility \cite{Fodor_Tc}) over $T_c(\chi_s)\approx
T_c(P)=176\pm3\pm4$\,MeV (extracted from the strange quark number 
and Polyakov loop susceptibilities \cite{Fodor_Tc}) to 
$T_c=192\pm7\pm4$\,MeV (extracted from light quark number and Polyakov
loop susceptibilities \cite{Karsch_Tc}). The differences between the
numbers extracted by different groups \cite{Fodor_Tc,Karsch_Tc} are
larger than the quoted statistical and systematic errors and even exceed
the FWHM widths $\Delta T_c \simeq 30-40$\,MeV of the various
susceptibilities \cite{Fodor_Tc}; they mostly reflect different 
choices for the observables used to set the physical scale and the
procedures employed for extrapolating to zero lattice spacing
\cite{Fodor_Tc,Karsch_Tc}.

From experimental data collected in $\sqrt{s}=200\,A$\,GeV Au+Au collisions 
at RHIC we know that the final hadron abundances from central collisions 
can be described by a hadron resonance gas in a state of approximate 
chemical equilibrium at $T_\mathrm{chem}=163\pm4$\,MeV, $\mu_B=24\pm4$\,MeV, 
and a strangeness saturation factor $\gamma_s=0.99\pm0.07$ \cite{STAR_Tchem}. 
The quality of the statistical model fit is impressive. Furthermore,
the STAR collaboration studied the dependence of the fit parameters
on the collision centrality and found that neither the temperature 
$T_\mathrm{chem}$ nor the baryon chemical potential $\mu_B$ depend 
appreciably on the impact parameter \cite{STAR_Tdec}\footnote{ The 
  chemical decoupling temperatures extracted from the measured hadron 
  abundance ratios depend somewhat on the details of the hadron resonance
  gas model employed: Ref.~\cite{STAR_Tchem} gives 
  $T_\mathrm{chem}=163\pm4$\,MeV, $\mu_B=24\pm4$\,MeV for central Au+Au
  collisions whereas in Ref.~\cite{STAR_Tdec} one finds 
  $T_\mathrm{chem}=157\pm6$\,MeV, $\mu_B=22\pm4$\,MeV for the same
  data. We will here use the values from the centrality dependence study 
  presented in Ref.~\cite{STAR_Tdec} which are consistently on the lower
  end of this range.}; only the 
strangeness suppression factor exhibits centrality dependence, beginning 
at impact parameters $>8-9$\,fm, and drops to values around 0.55 in the most
peripheral Au+Au collisions \cite{STAR_Tchem}. The centrality independence
of $T_\mathrm{chem}$ is in stark contrast to the behavior observed in
the same experiment for the kinetic decoupling temperature $T_\mathrm{kin}$
which is extracted (together with a value for the average radial flow 
velocity $\langle\beta\rangle$ of the fireball at kinetic freeze-out) 
from the shape of the transverse momentum spectra of identified pions,
kaons and (anti-)protons \cite{STAR_Tdec}: $T_\mathrm{kin}$ increases
significantly with increasing impact parameter, from 
$T_\mathrm{kin}=89\pm12$\,MeV in the most central to   
$T_\mathrm{kin}=127\pm13$\,MeV in the most peripheral collisions. At the
same time the average radial flow decreases from 
$\langle\beta\rangle=0.59\pm0.05$ in the most central to
$\langle\beta\rangle=0.24\pm0.08$ in the most peripheral Au+Au collisions,
demonstrating a strong centrality dependence of the fireball expansion
dynamics.

This characteristic difference in the centrality dependences of the chemical
and kinetic decoupling temperatures will be the main focus of this
contribution. We will show that the observed centrality dependences
of the average radial flow velocity and kinetic freeze-out temperature are
consistent with hydrodynamic behaviour of the fireball medium followed
by kinetic decoupling of the hadrons from microscopic scattering processes,
driven by the collective expansion. We will then show that a centrality
independent freeze-out temperature is inconsistent with a kinetic decoupling
process unless the chemical scattering rates have an extremely (namely almost 
infinitely) strong temperature dependence. We interpret this finding as
evidence that chemical decoupling of the hadron abundances is driven by a 
phase transition during which the chemical reaction rates decrease 
precipitously, leaving the system in a chemically frozen-out state at the 
end of the phase transition. Only in this way is it possible to obtain
a universal chemical freeze-out temperature that is insensitive to the 
(centrality dependent) collective dynamics and only depends on the 
thermodynamic parameters of the phase transition. Obviously, the chemical 
processes happening during the hadronization process itself involve colored 
degrees of freedom and can thus not be efficiently described in hadronic 
language. We also address the centrality dependence of the strangeness 
saturation factor and comment on how our picture also reproduces chemical 
abundance data measured in $pp$ and $e^+e^-$ collisions.  

\section{Review of proposed phase transition signatures}

In this section we will review several previously published suggestions
for QGP phase transition signatures in heavy-ion collisions and explain
why they either failed or their status has remained ambiguous. 

\subsection{Characteristics of the quark-hadron phase transition}

\begin{itemize}
\item
Lattice QCD tells us \cite{KL03} that the finite temperature, 
$\mu_B\approx 0$ QCD phase transition is associated with a rapid rise
of the normalized energy and entropy densities, $e/T^4$ and $s/T^3$,
near $T_c$, indicating a sudden change of the number of effective 
degrees of freedom. 
\item
The square of the speed of sound $c_s^2=\frac{\partial p}{\partial e}$ 
exhibits a deep minimum near $T_c$, dropping from close to the ideal gas 
value $c_s^2=\frac{1}{3}$ at $T\gtrsim 2\,T_c$ to almost 1/10 of that 
value near $T_c$ \cite{KL03} before rising again below $T_c$ to about 
half the ideal gas value, as calculated from the hadron resonance gas 
model \cite{QGP3}. (This is followed by a final exponential decrease 
as $T\to0$ which is, however, not phenomenologically relevant since 
the fireball matter freezes out before reaching such low temperatures). 
\item
Thermal fluctuations of the net 
electric charge and baryon number decrease by a factor 2-3 above $T_c$,
as a result of the charge and baryon number being distributed in smaller
(fractional) units. 
\item
The transport properties of the matter, such as
color conductivity $\sigma^{ab}$ \cite{AP2} and specific shear viscosity 
$\eta/s$ \cite{HG06}, change dramatically at $T_c$. 
\item
Finally, chemical 
equilibration rates, in particular for processes involving the creation and
destruction of strange quark pairs \cite{RM82}, speed up considerably above 
$T_c$, due to reduced mass thresholds as a result of chiral symmetry 
restoration.  
\end{itemize}

All these changes refer to thermal equilibrium properties of the matter
created in the collision, with the temperature as control parameter.
Unfortunately, near $T_c$ a few percent rise of the temperature requires
a severalfold increase of the energy density, so these changes set in much 
more slowly when viewed as a function of experimental conditions such as 
the collision energy, and it may be hard to recognize them among other 
effects. Finally, the medium produced in the collision cools through the 
phase transition quite rapidly, and even if it manages to remain near 
thermal equilibrium in spite of its strong collective dynamics the question 
remains which of these phase transition characteristics survive the 
expansion and affect the final hadronic state after freeze-out in a 
recognizable fashion.

\subsection{A visit to the graveyard of phase transition signatures}

The first attempt to connect the rapid rise of the entropy density
$s$ at $T_c$ with experimental data was made by van Hove \cite{VanHove82}
who suggested to plot the average transverse momentum $\langle p_T\rangle$
of final state hadrons (as a proxy for the temperature) against their
rapidity density $dN/dy$ (as a proxy for entropy density \cite{Hwa85}) -- 
the phase transition should then show up as a ``plateau'', i.e. as a 
(limited) region where $\langle p_T\rangle \sim T$ remains constant while 
the entropy density $s \sim (1/\pi R_A^2)(dN/dy)$ keeps rising with 
increasing collision energy. However, McLerran {\it et al.} soon
realized \cite{KLRG86} that the average transverse momentum 
$\langle p_T\rangle$ of the final hadrons reflects a combination of
random thermal and outwardly directed collective motion {\em at freeze-out}
and is thus not a good proxy for the initial fireball temperature.
Due to the collective flow component, $\langle p_T\rangle$ depends on the 
hadron rest mass, and since larger initial entropy or energy densities 
(generated by higher beam energies) increase the lifetime of the fireball 
before freeze-out, this collective flow component keeps increasing as the 
beam energy is raised even when the initial fireball temperature hardly 
changes inside the phase transition. As a result, there
is no ``plateau'' of $\langle p_T\rangle$ vs. $dN/dy$ -- the average
transverse momentum increases monotonically with $dN/dy$, due to the
monotonic increase of the collective radial flow with collision energy
\cite{KLRG86}. This growth is somewhat slowed in the phase transition 
region where the thermal $\langle p_T\rangle$ component remains constant, 
but that is difficult to observe, and it is even harder then to make a 
convincing case that such a change of slope in $\langle p_T\rangle$ vs. 
$dN/dy$ is indeed caused by a phase transition.

The speed of sound of the fireball medium represents its ``pushing power'',
i.e. its ability to accelerate its collective flow in response to pressure 
gradients. Near a phase transition it becomes small (or even vanishes if
the transition is of first order), thus it has been suggested early on
(especially by W.~Greiner, H.~St\"ocker and collaborators \cite{Ffm}) that
phase transitions should be visible in the collective flow pattern of the
matter created in heavy ion collisions. A specific suggestion made by 
Rischke et al. \cite{Rischke95} within a 1-fluid hydrodynamic model was
that the directed flow (``bounce-off''), when measured as a function of 
beam energy, should collapse around $E_\mathrm{lab}=5\,A$\,GeV due to
the softness of matter created near the phase coexistence region, but 
recover again at higher beam energies when the fireball is initially in 
the QGP phase. Subsequent studies within 3-fluid hydrodynamics \cite{B00}, 
where the two colliding nuclei and the matter created by their collision are
described as coupled but separately thermalized fluids, showed that this
phase transition signature is fragile and exhibits great sensitivity
to the non-equilibrium dynamics describing the transfer of energy between
the three fluids. In a realistic scenario the characteristic minimum in
the excitation function of the directed flow completely disappears \cite{B00}.

A variation of this theme was proposed in \cite{KSH00} whose authors
suggested that the softness of the Equation of State (EoS) near a phase
transition would manifest itself in a non-monotonic elliptic flow excitation
function, with a minimum at collision energies high enough to initialize
the medium slightly above the phase transition such that most of the elliptic
flow would develop while the matter cools through the softest point.
The advantage of this suggestion was that elliptic flow would be created
only after the matter has thermalized, thereby avoiding the very early
pre-equilibrium stage that proved fatal for the directed flow collapse.
However, the elliptic flow minimum turned out to be fragile, too, only 
this time the phase transition signal was killed by non-equilibrium effects 
during the late instead of the early collision stages \cite{TLS01,HHKLN}:
At beam energies below the predicted minimum, the elliptic flow would have
to be created in the hadron resonance gas phase (whose EoS is less stiff
than that of the QGP but still much stiffer than the matter near the
phase transition), but the latter is so viscous that its elliptic flow
response to anisotropic pressure gradients is dramatically reduced. As a result,
the elliptic flow signature never recovers at lower beam energies, thereby
wiping out the predicted flow minimum around RHIC energies.

Thus, even though there is now strong evidence from a large variety of 
measurements that the dense matter created in Au+Au collisions at RHIC is 
initially in the QGP phase \cite{PANIC02}, no direct evidence exists that 
it indeed passes through a phase transition on its way to the final state of 
frozen-out hadrons. Attempts to capitalize on the predicted change in
the net charge and baryon number fluctuation spectrum \cite{AHM00}
have not paid off -- it seems that the predicted reduction in event-by-event 
fluctuations is largely washed out by the hadronization process. 

Much attention has been paid to non-monotonicities in the excitation 
functions of several heavy-ion observables measured by the NA49 
Collaboration in Pb+Pb collisions at low SPS energies: a ``kink'' in the pion 
production per wounded nucleon, a ``step'' in the slope parameter of the 
$K^+$ transverse momentum spectrum, and a ``peak'' in the ratio of produced 
strange to non-strange quark-antiquark pairs (for a review see 
Ref.~\cite{Sey}). The suggested interpretation of these observations in 
terms of a phase transition \cite{Gaz,Sey} is controversial since it relies 
on a number of unverivied simplifying assumptions and is so far not backed up
by dynamical model calculations. 
  
All the above suggestions are based on measuring excitation functions
in order to steer the medium in a controlled way through the phase 
transition. The extraction of $T_c$ from such studies requires a model 
to quantitatively relate the control parameters $\sqrt{s}$ or $dN/dy$ to 
the initial energy density (and thus, via the assumption of thermalization), 
to a temperature).

We will here propose a different approach that exploits the centrality
dependence of observables at fixed collision energies. While it has been
pointed out repeatedly that the initial energy density of the created 
matter can be varied either by changing the collision energy for fixed
collision system or by changing the size of the colliding nuclei or their
impact parameter at fixed collision energy, we are not relying primarily
on a controlled change of the initial energy density, but on the associated
variation of the collective expansion rate. In order to explain how we use
this, let us first discuss the recent controversy relating to the 
interpretation of chemical freeze-out data from relativistic heavy-ion 
collisions which motivated our work.  

\subsection{The controversy: Chemical reaction kinetics vs. statistical 
hadronization}

The hadrons emitted in relativistic heavy-ion collisions show thermal
characteristics both in their abundances and in the shapes of their
transverse momentum spectra, but the temperatures extracted from
yields \cite{STAR_Tchem,STAR_Tdec,CR98,BRS03} (``chemical freeze-out 
temperature'' $T_\mathrm{chem}$) and from spectra 
\cite{STAR_Tdec,Dobler,CR99,Tomasik,JBH03} (``kinetic decoupling 
temperature'' $T_\mathrm{kin}$) differ significantly. The kinetic
decoupling temperatures depend on beam energy and, at all beam energies,
on system size and collision centrality, while for $\sqrt{s}\gtrsim 
10\,A$\,GeV approximately the same ``universal'' chemical freeze-out
temperature is observed in $e^+e^-$, $p\bar p$ and $A+A$ collisions at
all collision centralities (only the strangeness saturation factor 
$\gamma_s$ varies with collision system and centrality) \cite{Becattini}. 

This universal value $T_\mathrm{chem}=160-170$\,MeV is remarkably close 
to the critical temperature for the quark-hadron transition from Lattice 
QCD. Furthermore, kinetic simulations of the hadronic rescattering stage
after QGP hadronization with hadronic cascade codes (RQMD, UrQMD, etc.)
have shown that hadronic rescattering alters the momentum distributions 
and resonance populations through resonance scattering, thereby cooling 
the system while keeping it (at least for a while) close to local thermal 
equilibrium \cite{Bravina}, while leaving the final stable hadron yields 
(after resonance decays) almost unaffected \cite{BD99,H99}. Hadronic 
rescattering leads to the loss of a fraction of the baryon-antibaryon 
pairs, but this can be at least partially traced back to the absence
of multi-hadron collision channels so that detailed balance is violated
in baryon-antibaryon annihilation channels \cite{RS01,GL01}.
 
These empirical facts have split the heavy-ion theory community into
two camps which offer different interpretations of the observations.
The philosophy of {\bf Camp I} is laid out in Refs.~\cite{Becattini} 
(second paper) and \cite{H98,H99,S99} and holds that hadron production
is a statistical process associated with a phase transition, proceeding
through very many different possible microscopic channels constrained
only by energy, baryon number and strangeness conservation, thereby leading
to a maximum entropy configuration described by a thermal distribution
of hadron yields, with $T_\mathrm{chem}$, $\mu_B$ and $\gamma_s$ playing 
the role of Lagrange multipliers to ensure these conservation law constraints
while maximizing the entropy. The value of $T_\mathrm{chem}$ is {\em not} 
established by inelastic reactions among hadrons proceeding until
chemical equilibrium is reached  -- rather, the hadrons are directly ``born'' 
into a maximum entropy state of apparent chemical equilibrium \cite{S99}, 
with the parameter $T_\mathrm{chem}$ defining the critical energy density 
at which the hadronization process happens (\cite{Becattini}b). 
$T_\mathrm{chem}$ is thus conceptually different from the kinetic decoupling 
temperature $T_\mathrm{kin}$ which {\em is} the result of quasi-elastic 
rescattering among the hadrons (which also contribute to their collective 
flow).  

{\bf Camp II} includes the followers of Refs.~\cite{RS01,GL01,BSW04,GrSQM04}
who hold that chemical freeze-out is a kinetic process within the hadronic
phase, conceptually equivalent with kinetic freeze-out, the only difference
being the quantitative values of the corresponding freeze-out temperatures
which reflect the fact that the inelastic cross sections driving chemical 
equilibration constitute only a small fraction of the total scattering
cross section contributing to momentum exchange. The hadrons are not born
into chemical equilibrium, but driven into such a state kinetically by
inelastic multi-hadron processes (which, according to 
Refs.~\cite{RS01,GL01,BSW04,GrSQM04}, become crucial near $T_c$ due to 
high hadron densities) and frozen out by global expansion. Accordingly,
$T_\mathrm{chem}$ is the ``real'' temperature at which forward and backward 
chemical reactions last balance each other. (In contrast, for Camp I, there
are no ``backward'' reactions involving hadrons in both initial and final 
states.)

Is this more than a philosophical difference of opinions? We think so. 
Camp II has 
to cope with an intrinsic tension between two observations: The high 
quality of the thermal model fit to the observed hadron yields at RHIC 
requires sufficient time for inelastic reactions to establish chemical 
equilibrium, whereas the proximity of the fitted chemical freeze-out 
temperature $T_\mathrm{chem}$ to the critical temperature $T_c$ of the 
quark-hadron phase transition from Lattice QCD, together with the rapid 
cooling of the fireball by collective expansion, don't provide much of a 
window for these processes to play out. In essence, to make the kinetic 
chemical equilibration scenario work one needs {\em very} large scattering 
rates right near $T_c$ which then drop to negligible values just below 
$T_c$. This would be easier to understand if there were a larger gap between 
$T_c$ and $T_\mathrm{chem}$, as suggested by the recent upward revision 
of $T_c$ from Lattice QCD advocated in \cite{Karsch_Tc}, but this problem
appears serious if the lower $T_c(\chi_{\bar\psi\psi})$ from 
Ref.~\cite{Fodor_Tc} turns out to be correct. 

In addition, we point out another conceptual problem with the 
interpretation by Camp II: If freeze-out is a kinetic process, it
is controlled by the competition between local scattering (moving
the system towards equilibrium) and global expansion (driving the 
system out of equilibrium). The resulting freeze-out temperature
is therefore sensitive to the fireball expansion rate. We show that 
the latter depends on collision centrality. Therefore, any kinetic 
freeze-out temperature must depend on impact parameter. While this is 
empirically indeed the case for the kinetic decoupling temperature
$T_\mathrm{kin}$, the chemical freeze-out temperature does not seem
to vary with collision centrality. Hence it cannot be the result of a
kinetic decoupling process from inelastic hadronic scattering.

\section{Kinetic freeze-out from a hydrodynamically expanding system}

In this section we show that the hydrodynamic model can quantitatively 
reproduce the observed centrality dependence of the kinetic decoupling
temperature extracted from hadron momentum spectra at RHIC. We then show
that an analogous centrality dependence of the chemical freeze-out 
temperature cannot be avoided if the hadron yields are similarly controlled 
by kinetic freeze-out from inelastic hadronic rescattering. 

We use our (2+1)-dimensional longitudinally boost-invariant hydrodynamic
AZHYDRO \cite{AZHYDRO} with standard initial conditions \cite{QGP3} to 
generate the flow pattern for $200\,A$\,GeV Au+Au collisions. This code
has been previously shown to successfully reproduce the measured single 
particle hadron $p_T$-spectra and their elliptic flow (for details see
\cite{H_SQM04}). Here, however, we modify the freeze-out criterium for 
kinetic decoupling to account for its kinetic nature: Instead of requiring
freeze-out on a surface of constant energy density $e_\mathrm{dec} =
0.075$\,GeV/fm$^3$ (corresponding to a fixed temperature 
$T_\mathrm{kin}=100$\,MeV \cite{QGP3}), we define the kinetic freeze-out 
surface as the set of points where the local expansion rate 
$1/\tau_\mathrm{exp}(x)= \partial\cdot u(x)$ ($u^\mu(x)$ being the
hydrodynamic flow 4-velocity) becomes equal to the local scattering rate 
$1/\tau_\mathrm{scatt}(x)$ \cite{BGZ78,Schnekki,Hung}:  
\begin{equation}
\label{focrit}
  \frac{1}{\tau_\mathrm{scatt}} = \xi\, \frac{1}{\tau_\mathrm{exp}} 
                                = \xi\, \partial\cdot u(x).
\end{equation}
Here $\xi$ is a proportionality constant of order unity which we set in 
a first attempt to $\xi=0.35$, yielding an average temperature along the 
freeze-out surface for central Au+Au collisions of 
$\langle T_\mathrm{kin}\rangle\simeq 115$\,MeV. Having fixed $\xi$ in 
central collisions, Eq.~(\ref{focrit}) is taken to define the freeze-out 
surface also at other impact parameters. Inside the freeze-out surface 
the scattering rate exceeds $\xi$ times the expansion rate, and the matter 
is thermalized, whereas outside the surface the expansion rate exceeds 
$\xi^{-1}$ times the scattering rate so we assume the hadrons there to 
be decoupled from the fluid, streaming freely into the detector.

The expansion rate $\partial\cdot u=\gamma_\perp\left(\frac{1}{\tau}
+\bm{\nabla}\cdot\bm{v}_\perp\right) + \left(\partial_\tau + 
\bm{v}_\perp{\cdot}\bm{\nabla}_\perp\right) \gamma_\perp$ is computed
from the hydrodynamic output for the transverse flow velocity 
$\bm{v}_\perp(x)$ and $\gamma_\perp=(1+v_\perp^2)^{-1/2}$. 

The scattering rate in Eq.~(\ref{focrit}) involves cross sections and 
densities of scatterers, hence it is particle specific \cite{Schnekki,Hung}.
Since at RHIC energies hadron production is dominated by pions, we assume
for simplicity that all hadrons decouple when pions freeze out. It may be
somewhat better to describe different hadron species by different freeze-out 
temperatures \cite{Hung}, but each of them would exhibit similar impact 
parameter dependences as the one for pions which we work out here.

The pion scattering rate we take from the numerical results presented
in Ref.~\cite{Hung} which we parametrize as
\begin{equation}
\label{rate}
  \frac{1}{\tau^\pi_\mathrm{scatt}} = \left(59.5\,\mathrm{fm}^{-1}\right)
  \left(\frac{T}{1\,\mathrm{GeV}}\right)^{3.45}.
\end{equation}
This defines the momentum exchange rate to be used for the calculation 
of kinetic freeze-out and will need to be modified when discussing
chemical freeze-out below.

%
%%%%%%%%%%%%%%%%%%%%%%%%%%%%%%%%%%%%% Fig. 1 %%%%%%%%%%%%%%%%%%%%%%%%%%%%%%%
\begin{figure}[ht]
%\centerline{\epsfig{file=dndm2_0.ps,bb=51  208  554  575,width=7.9cm}}
\centerline{\epsfig{file=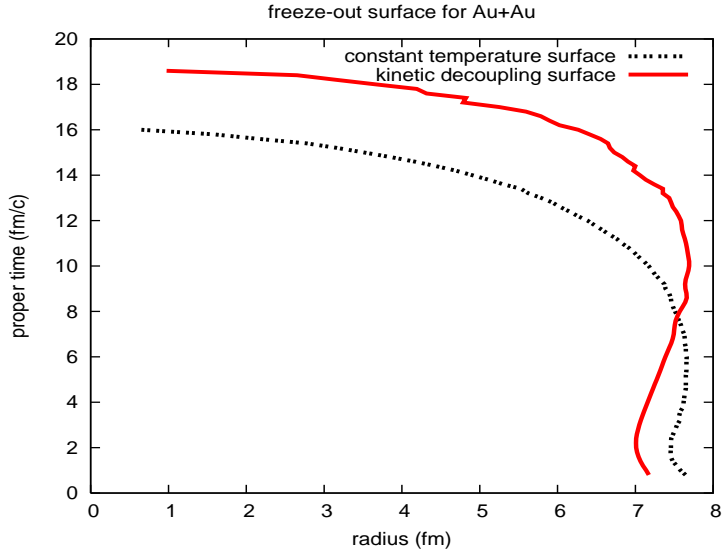,width=10cm,height=7.5cm}}
\caption{Kinetic freeze-out surface $\tau_\mathrm{kin}(r)$ for central 
($b=0$) $200\,A$\,GeV Au+Au collisions, computed from 
Eq.~(\protect\ref{focrit}) 
with $\xi=0.35$ (red solid line) and for a constant freeze-out temperature
$T_\mathrm{kin}=115$\,MeV (dotted black line). Both surfaces have the
same average temperature of $\langle T\rangle =115$\,MeV, using the 
energy density as weight function.
\label{F1}}
\end{figure}
%%%%%%%%%%%%%%%%%%%%%%%%%%%%%%%%%%%%%%%%%%%%%%%%%%%%%%%%%%%%%%%%%%%%%%%%%%%%
%

In Fig.~\ref{F1} we show the kinetic freeze-out surface for central Au+Au
collisions computed from Eq.~(\ref{focrit}) with $\xi=0.35$ (solid red line)
and from the condition $T_\mathrm{kin}=115$\,MeV (dotted black line). Both
have the same average kinetic freeze-out temperature, but for the kinetic 
freeze-out criterium (\ref{focrit}) the middle of the fireball freezes out
a bit later at lower temperature and larger flow whereas the edge decouples
earlier at higher temperature and with less flow than the contant-$T$ 
surface. This is caused by the larger expansion \emph{rate} near the
edge of the fireball.

%
%%%%%%%%%%%%%%%%%%%%%%%%%%%%%%%%%%%%% Fig. 2 %%%%%%%%%%%%%%%%%%%%%%%%%%%%%%%
\begin{figure}[ht]
%\centerline{\epsfig{file=dndm2_0.ps,bb=51  208  554  575,width=7.9cm}}
\centerline{\epsfig{file=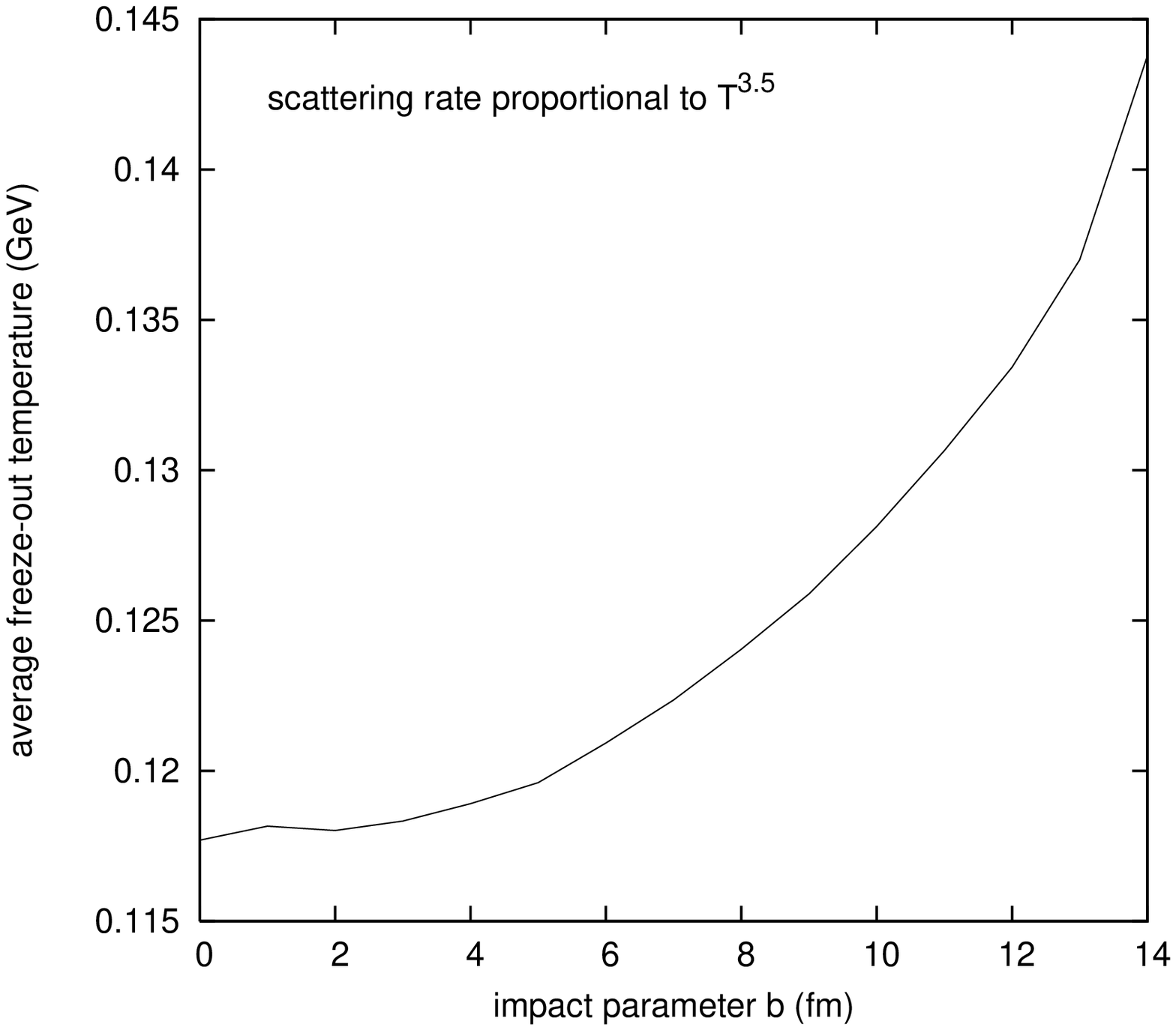,width=7.6cm,height=5cm}
            \epsfig{file=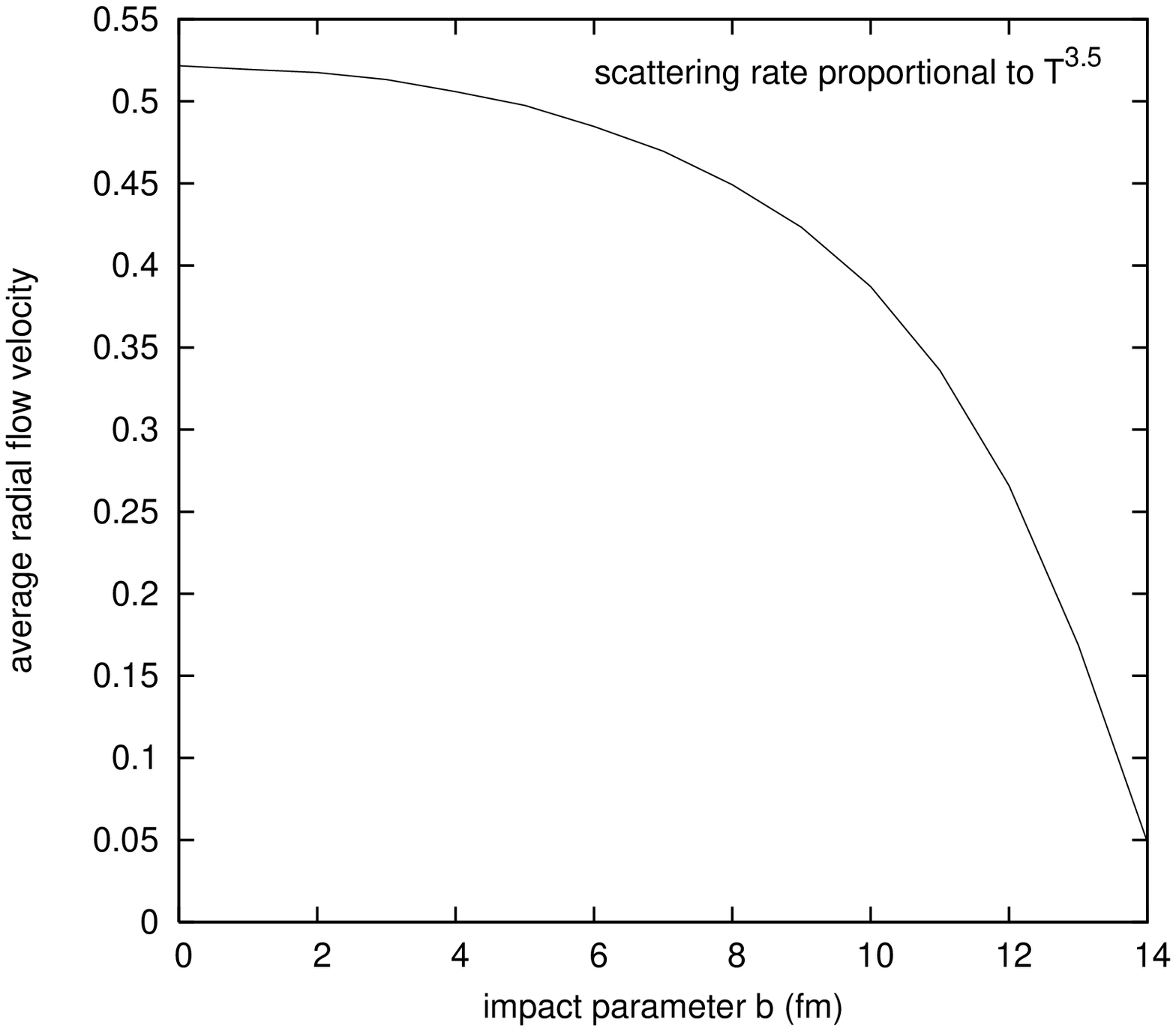,width=7.7cm,height=5cm}}
\caption{Impact parameter dependence of the average kinetic decoupling
temperature $\langle T_\mathrm{kin}\rangle$ (left) and average radial 
flow velocity $\langle v_\perp\rangle$ along the freeze-out surface
computed from Eq.~(\protect\ref{focrit}) with $\xi=0.35$ and a scattering 
rate $\sim T^{3.45}$ as in Eq.~(\protect\ref{rate}), for $200\,A$\,GeV
Au+Au collisions.
\label{F2}}
\end{figure}
%%%%%%%%%%%%%%%%%%%%%%%%%%%%%%%%%%%%%%%%%%%%%%%%%%%%%%%%%%%%%%%%%%%%%%%%%%%%
%

Figure~\ref{F2} shows the impact parameter dependence of the average kinetic
decoupling temperature and the associated average radial flow calculated
from the kinetic freeze-out criterium (\ref{focrit}) with $\xi=0.35$. One 
sees that central collisions decouple at relatively low temperatures with
large average radial flow whereas peripheral collisions freeze out earlier
when the fireballs are still hotter and less radial flow has developed.
This is in good qualitative agreement with the STAR data in
Ref.~\cite{STAR_Tdec}, although there the freeze-out temperatures are 
generally a bit lower, with slightly larger average radial flow velocities
than seen in Fig.~\ref{F2}.
%
%%%%%%%%%%%%%%%%%%%%%%%%%%%%%%%%%%%%% Fig. 3 %%%%%%%%%%%%%%%%%%%%%%%%%%%%%%%
\begin{figure}[ht]
%\centerline{\epsfig{file=dndm2_0.ps,bb=51  208  554  575,width=7.9cm}}
\centerline{\epsfig{file=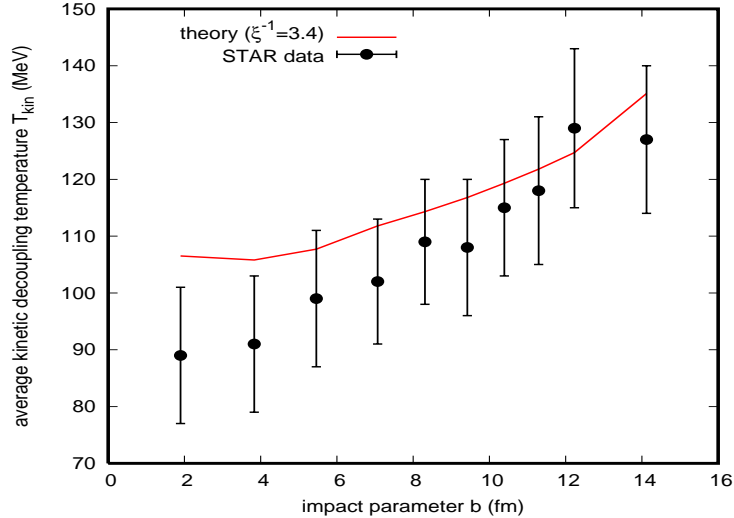,width=10cm,height=7cm}}
\caption{Impact parameter dependence of the average kinetic decoupling
temperature $\langle T_\mathrm{kin}\rangle$ computed from hydrodynamics with
kinetic freeze-out criterium (\protect\ref{focrit}) using $\xi=0.295$, 
compared with STAR data \protect\cite{STAR_Tdec} for $200\,A$\,GeV
Au+Au collisions.
\label{F3}}
\end{figure}
%%%%%%%%%%%%%%%%%%%%%%%%%%%%%%%%%%%%%%%%%%%%%%%%%%%%%%%%%%%%%%%%%%%%%%%%%%%%
%
We adjust for this by fine-tuning the phenomenological parameter $\xi$
in Eq.~(\ref{focrit}) to $\xi=0.295$ ($\xi^{-1}=3.4$). The corresponding
freeze-out temperatures are shown as a function of impact parameter $b$
in Figure~\ref{F3}, together with the STAR data. Now the agreement is also
quantitatively acceptable. We conclude that the measured centrality dependence
of $T_\mathrm{kin}$ can be completely understood in terms of a hydrodynamic
model for the fireball expansion coupled to a kinetic freeze-out criterium
with realistic temperature dependence of the microscopic scattering rate.  
	
Let us now see whether we can similarly understand chemical freeze-out
as a kinetic decoupling process from inelastic hadronic scattering. A
few typical processes relevant for chemical equilibration are
\begin{eqnarray}
\label{inel}
&&\pi + \pi \longleftrightarrow K + \bar K, \qquad
  \pi + N \longleftrightarrow K + Y,, \qquad
  \pi + Y \longleftrightarrow \bar K + N,  
\nonumber\\
&&\Omega + \pi \longleftrightarrow \Xi + \bar K,\qquad
  K + \bar K \longleftrightarrow \phi + \pi,\qquad
  \Omega + \bar K \longleftrightarrow \Xi + \pi,
\\
&&\Omega + \bar N \longleftrightarrow 2 \pi + 3\bar K,\qquad
  N + \bar N \longleftrightarrow 5\pi,\qquad
  N + 3\bar K \longleftrightarrow \Omega + 3\pi.
%  \Xi + \bar Y \longleftrightarrow 4\pi + \bar K.
\nonumber
\end{eqnarray}
The last line shows so-called multi-hadron collision channels which in at 
least one direction require collisions between more than two hadrons.
Rates for processes involving $n_\mathrm{in}$ incoming hadrons are
proportional to the product of their densities $\sim \Pi_{i=1}^{n_\mathrm{in}}
n_i(T)$ where each factor $n_i(T)$ grows with $T$ at least as $T^3$ (even 
much more rapidly for hadrons with masses $>T$). At low temperatures, 
multi-hadron collision processes as well as collisions between very massive 
hadrons are therefore strongly suppressed. Consequently, particle yields for 
hadrons requiring collisions of many abundantly available particles for their 
production or destruction (such as $\bar p, \Omega, \dots$) thus tend to 
freeze out at higher $T$ than particle yields for hadrons whose abundances 
can be efficiently changed by two-body reactions ($\pi, K, \phi, \dots$).

In an expanding, cooling system, simultaneous freeze-out of all hadron 
yields at a {\em common} temperature therefore requires a {\em conspiracy 
of rates} with widely differring $T$-dependences. Indeed, thermal model fits 
to hadron abundances with a single common temperature are usually not
perfect \cite{SHSX98}, and individual fits to subsets of yields measured
in lower-energy collisions at the SPS and AGS tend to lead to a significant 
spread of chemical freeze-out temperatures \cite{DPZ06}. Only at RHIC
energies so far the single-temperature chemical equilibrium fit gives
an almost perfect fit to the data \cite{BRS03,DPZ06}.

One way to achieve the conspiracy of different chemical equilibration 
rates that is required for a good fit with a single freeze-out temperature 
is to postulate that at chemical freeze-out {\em all} chemical reactions
are completely dominated by multi-hadron collisions and that at any
temperature below $T_\mathrm{chem}$ the medium is so rarefied and so 
rapidly expanding that even the simplest two-body reactions among the
most abundantly produced hadrons (such as those listed in the first line
of Eq.~(\ref{inel})) have essentially stopped. As long as collision 
channels with widely different temperature dependences compete with
each other, chemical freeze-out of all hadron species at a single 
temperature appears to impossible.

\begin{figure}[ht]
%\centerline{\epsfig{file=dndm2_0.ps,bb=51  208  554  575,width=7.9cm}}
\centerline{\epsfig{file=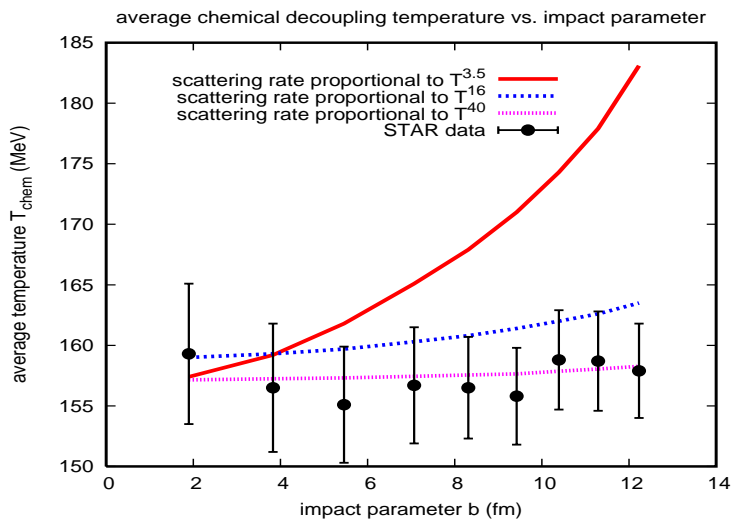,width=10cm,height=7cm}}
\caption{Impact parameter dependence of the average chemical decoupling
temperature $\langle T_\mathrm{chem}\rangle$ computed from hydrodynamics 
with kinetic freeze-out criterium (\protect\ref{focrit}) using $\xi=0.95$
and reaction rates with different temperature dependences as listed, 
compared with STAR data \protect\cite{STAR_Tdec} for $200\,A$\,GeV
Au+Au collisions.
\label{F4}}
\end{figure}

Even more importantly, even if it were possible at one fixed impact 
parameter to arrange for common freeze-out of all hadron species in 
spite of a competition of scattering rates with different temperature
dependences, such a conspiracy would be impossible to maintain, {\em 
with the same value for the freeze-out temperature $T_\mathrm{chem}$},
over the entire impact parameter range. Figure~\ref{F4} shows the 
%
%%%%%%%%%%%%%%%%%%%%%%%%%%%%%%%%%%%%% Fig. 4 %%%%%%%%%%%%%%%%%%%%%%%%%%%%%%%
%%%%%%%%%%%%%%%%%%%%%%%%%%%%%%%%%%%%%%%%%%%%%%%%%%%%%%%%%%%%%%%%%%%%%%%%%%%%
%
centrality dependence of the average chemical freeze-out temperatures 
along hydrodynamic decoupling surfaces computed with the kinetic 
freeze-out criterium (\ref{focrit}), using $\xi=0.95$ to adjust the 
value of $\langle T_\mathrm{chem}\rangle$ in central Au+Au collisions 
to the STAR data \cite{STAR_Tdec} and exploring different possible 
temperature dependences of the dominant inelastic scattering rate. One 
sees that approximate impact parameter independence of 
$\langle T_\mathrm{chem}\rangle$ can only be achieved if \emph{all} 
inelastic scattering rates grow with $T$ as $T^n$ with a power 
$n\gtrsim20$!

Basically Figure~\ref{F4} tells us that the observed centrality
independence of $T_\mathrm{chem}$ requires chemical freeze-out to
happen in a region of parameter space where {\em all} chemical 
reaction rates exhibit extremely steep temperature dependence, dropping
like a stone as the system cools through the decoupling temperature.
It is hard to understand such a behavior within a hadron rescattering 
picture unless one assumes that {\em all} relevant chemical reactions 
involve multi-particle channels involving many hadrons. Making such
an assumption clearly pushes the hadronic rescattering model towards 
breakdown because its chemical kinetics would essentially be controlled 
by interactions among clusters of particles involving an unspecifiable 
number of hadrons. {\em It is much more natural to associate this 
kind of behavior with the quark-hadron phase transition} where densely 
spaced and strongly interacting quarks and gluons provide the necessary 
multi-particle clusters, and where the dramatic change in number and 
quality of the effective degrees of freedom within a narrow temperature 
interval generates the dramatic temperature dependence of the chemical 
reaction rates at decoupling which seem to be phenomenologically 
required.  

In such a picture, hadrons are not really well-defined states until after
the quark-hadron phase transition is complete and, at the same time, 
chemical reactions among hadrons have ceased. Hadrons are thus indeed
``born into chemical equilibrium'' \cite{S99} in a process that can be
rightfully called ``statistical hadronization'' \cite{Becattini,H98,H99}.
If hadrons are formed in this fashion, their measured abundances provide 
a window with a direct view of the QCD quark-hadron phase transition. 

\section{Conclusions}
 
We have shown that the observed impact parameter dependence of the average
temperature and radial flow velocity at {\em kinetic (thermal) freeze-out}
(i.e. at the point where the hadron momentum distributions decouple) can be 
quantitatively understood as a kinetic decoupling process in a 
hydrodynamically expanding source, with freeze-out being driven by the
global expansion of the collision fireball. Any such kinetic decoupling
process is controlled by the local competition between temperature 
dependent scattering and hydrodynamic expansion rates, and since the latter
change with impact parameter as a result of the varying initial energy 
density and size of the nuclear collision zone, the resulting average
freeze-out temperature is necessarily impact parameter dependent. The
strength of this impact parameter dependence (i.e. the sensitivity of
the freeze-out temperature to the fireball expansion rate) is inversely
related to the strength of the temperature dependence of the local 
scattering rate. To obtain approximate centrality independence of
the freeze-out temperature, the scattering rate must exhibit an almost 
infinitely steep temperature dependence.  

From this it follows that the observed impact parameter independence of 
the chemical freeze-out temperature in Au+Au collisions at RHIC (i.e. of 
the temperature where the abundances of stable hadron species decouple) 
cannot be consistently described as the result of a kinetic decoupling 
process from inelastic hadronic interactions. To obtain the necessary 
extremely steep temperature dependence of the inelastic scattering rate 
($\sim T^n$ with $n\gtrsim20$) requires that at the freeze-out point 
{\em all} chemical reactions are dominated by multi-hadron interactions 
involving many more than two colliding particles, in which case it seems 
unlikely that one will ever be able to describe this process quantitatively 
in hadronic language.  

In our opinion the only theoretically consistent interpretation of the STAR
data on chemical freeze-out is to associate the steepness of the the
temperature dependence of chemical equilibration rates with a phase
transition (in this case the quark-hadron transition) in which the
hadrons are produced statistically and distributed among different species
according to the principle of maximum entropy, via a multitude of 
complicated microscopic channels involving large numbers of strongly 
interacting quarks and gluons. In this sense the hadrons are ``born into
chemical equilibrium'' in an environment that is too dilute and expands too
rapidly to allow for {\em any} further inelastic reactions among the hadrons.

$T_\mathrm{kin}$ and $T_\mathrm{chem}$ thus stand on conceptually 
different footings. $T_\mathrm{chem}$ is a Lagrange multiplier related 
by the Maximum Entropy Principle to the critical energy density 
$e_c$ for hadronization. Its universality in $e^+e^-$, $pp$, and $AA$ 
collisions of all centralities shows that at $e_c$ a phase transition 
occurs. Hadrons are formed during this transition in a statistical 
process subject to the Principle of Maximum Entropy.

The absence of inelastic {\em hadronic} rescattering processes allows the 
direct measurement of $T_c$ through $T_\mathrm{chem}$ and thus the 
experimental observation of the phase transition. In this context the
question arises which of the different definitions of the critical
temperature $T_c$ from Lattice QCD that were mentioned in the Introduction
is most closely related to the chemical freeze-out temperature 
$T_\mathrm{chem}$ extracted from hadron yield data. It seems unlikely
that hadron yields can be considered frozen out before the hadrons have
more or less recovered their full vacuum masses, and this is related
to the restoration of the chiral condensate $\langle\bar\psi\psi\rangle$
to its vacuum value. We therefore suggest that $T_c(\chi_{\bar\psi\psi})$
\cite{Fodor_Tc} should be the LQCD number most closely related to the 
phenomenological value $T_\mathrm{chem}$. This seems to be consistent
with the actual values extracted in \cite{Fodor_Tc} from LQCD and in
\cite{STAR_Tdec} from hadron yields at RHIC (both are between 150 and 
160 MeV).

The increase of the strangeness saturation factor $\gamma_s$ from $e^+e^-$ 
and $pp$ to heavy-ion collisions and from peripheral to central Au+Au 
collisions at RHIC shows that the lifetime of the QGP (and thus the time 
for chemically equilibrating strange with light quarks) is still limited. 
Only for midcentral to central Au+Au collisions $\gamma_s$ has sufficient 
time to saturate. (Qualitatively similar tendencies are seen in Pb+Pb 
collisions at lower SPS energies \cite{BRS03}.) The primary parton 
production process at the beginning
of the collision apparently suppresses the production of strange quarks,
and it also produces $s$ and $\bar s$ locally in pairs, thereby generating
spatial correlations among $s$ and $\bar s$ which ensure strangeness 
conservation {\em locally}. In a grand canonical description such 
correlations induce a strangeness suppression factor $\gamma_s<1$ 
\cite{Becattini}. It takes time to diffuse the strange quarks over the
entire fireball volume to decorrelate them and adjust their abundance
to equilibrium values. Larger initial energy densities in central Au+Au
collision provide more time until the point of hadronization at 
$e_c\simeq0.7$\,GeV/fm$^3$ is reached than peripheral Au+Au or $e^+e^-$
and $pp$ collisions. 

We close by pointing out that our conclusions about the nature and origin of 
$T_\mathrm{chem}$ can be put to a relatively easy experimental test: It is
well known \cite{CR98,BRS03} that at low SPS and AGS energies, where the
net baryon density of the matter created in the collision is much larger
than at RHIC, the measured chemical decoupling temperatures are well below 
generally accepted estimates for the phase transition temperature,
$T_\mathrm{chem}<T_c$. In that case the phase transition can not be the 
origin of the observation of chemical equilibrium yields; hadronic chemical 
reactions must be responsible for lowering the chemical freeze-out 
temperature to values significantly below $T_c$. Since the present work 
has shown that the kinetic decoupling of hadronic chemical reaction rates 
is influenced by the fireball expansion rate, which again depends on 
collision centrality, {\em we expect to see impact parameter dependence of 
$T_\mathrm{chem}$ whenever its value is measured to be well below $T_c$}.
This conclusion would also apply to RHIC collisions if Lattice QCD would 
eventually converge to $T_c$ values above 190 MeV as proposed in 
\cite{Karsch_Tc}. In this case we would definitely expect $T_\mathrm{chem}$
to depend on collision centrality and ask for a reanalyzis of chemical
decoupling data at RHIC with higher statistics in order to unambiguously
settle this question.

It may be possible to re-analyze existing SPS data to confirm or falsify 
our prediction of centrality dependence of $T_\mathrm{chem}$ at these 
energies. If not, this will be a worthwhile point to address within the 
planned low-energy collision program at RHIC. Clarification of this point 
will be of utmost importance for establishing the observed chemical 
decoupling temperature at RHIC as a direct measurement of the critical 
temperature of the quark-hadron phase transition in QCD.

\end{document}